\documentclass[twocolumn,english,aps,preprintnumbers,amsmath,amssymb,superscriptaddress]{revtex4}
\usepackage[latin9]{inputenc}
\usepackage{amsmath}
\usepackage{graphicx}

\makeatletter
\@ifundefined{textcolor}{}
{%
 \definecolor{BLACK}{gray}{0}
 \definecolor{WHITE}{gray}{1}
 \definecolor{RED}{rgb}{1,0,0}
 \definecolor{GREEN}{rgb}{0,1,0}
 \definecolor{BLUE}{rgb}{0,0,1}
 \definecolor{CYAN}{cmyk}{1,0,0,0}
 \definecolor{MAGENTA}{cmyk}{0,1,0,0}
 \definecolor{YELLOW}{cmyk}{0,0,1,0}
}

\usepackage{dcolumn}% Align f columns on decimal point
\usepackage{bm}% bold math
\usepackage{color}
\usepackage[final]{pdfpages}

\usepackage{graphicx}
\usepackage[mathlines]{lineno}
\usepackage[svgnames]{xcolor}
\usepackage{hyperref}
\usepackage{color}

\usepackage{comment}
\usepackage{xspace}

\usepackage[version=4]{mhchem}
\usepackage{siunitx}
\usepackage{booktabs}

\newcommand{\rvb}{\ce{RbV3Sb5}\xspace}
\newcommand{\msr}{$\mu$SR\xspace}

\makeatother

\usepackage{babel}

\begin{document}

\preprint{preprint(\today)}

\title{Depth-dependent study of time-reversal symmetry-breaking in the kagome superconductor $A$V$_{3}$Sb$_{5}$}

\author{J.N.~Graham}
\thanks{These authors contributed equally to the paper.}
\affiliation{Laboratory for Muon Spin Spectroscopy, Paul Scherrer Institute, CH-5232 Villigen PSI, Switzerland}

\author{C.~Mielke III}
\thanks{These authors contributed equally to the paper.}
\affiliation{Laboratory for Muon Spin Spectroscopy, Paul Scherrer Institute, CH-5232 Villigen PSI, Switzerland}

\author{D.~Das}
\thanks{These authors contributed equally to the paper.}
\affiliation{Laboratory for Muon Spin Spectroscopy, Paul Scherrer Institute, CH-5232 Villigen PSI, Switzerland}

\author{T.~Morresi}
\affiliation{Dipartimento di Scienze Matematiche, Fisiche e Informatiche, Universit(\`a) di Parma, I-43124 Parma, Italy}

\author{V.~Sazgari}
\affiliation{Laboratory for Muon Spin Spectroscopy, Paul Scherrer Institute, CH-5232 Villigen PSI, Switzerland}

\author{A.~Suter}
\affiliation{Laboratory for Muon Spin Spectroscopy, Paul Scherrer Institute, CH-5232 Villigen PSI, Switzerland}

\author{T.~Prokscha}
\affiliation{Laboratory for Muon Spin Spectroscopy, Paul Scherrer Institute, CH-5232 Villigen PSI, Switzerland}

\author{H.~Deng}
\affiliation{Department of Physics, Southern University of Science and Technology, Shenzhen, Guangdong, 518055, China}

\author{R.~Khasanov}
\affiliation{Laboratory for Muon Spin Spectroscopy, Paul Scherrer Institute, CH-5232 Villigen PSI, Switzerland}

\author{S.D.~Wilson}
\affiliation{Materials Department, Materials Research Laboratory, and California NanoSystems Institute, University of California Santa Barbara, Santa Barbara, California 93106, USA}

\author{A.C.~Salinas}
\affiliation{Materials Department, Materials Research Laboratory, and California NanoSystems Institute, University of California Santa Barbara, Santa Barbara, California 93106, USA}

\author{M.M.~Martins}
\affiliation{Laboratory for Muon Spin Spectroscopy, Paul Scherrer Institute, CH-5232 Villigen PSI, Switzerland}

\author{Y.~Zhong}
\affiliation{Institute for Solid States Physics, The University of Tokyo, Kashiwa, Japan}

\author{K.~Okazaki}
\affiliation{Institute for Solid States Physics, The University of Tokyo, Kashiwa, Japan}

\author{Z.~Wang}
\affiliation{Centre for Quantum Physics, Key Laboratory of Advanced Optoelectronic Quantum Architecture and Measurement (MOE), School of Physics, Beijing Institute of Technology, Beijing, China}

\author{M.Z. Hasan}
\affiliation{Laboratory for Topological Quantum Matter and Advanced Spectroscopy (B7), Department of Physics,
Princeton University, Princeton, New Jersey 08544, USA}

\author{M. Fischer}
\affiliation{Physik-Institut, Universit\"{a}t Z\"{u}rich, Winterthurerstrasse 190, CH-8057 Z\"{u}rich, Switzerland}

\author{T. Neupert}
\affiliation{Physik-Institut, Universit\"{a}t Z\"{u}rich, Winterthurerstrasse 190, CH-8057 Z\"{u}rich, Switzerland}

\author{J.-X.~Yin}
\affiliation{Department of Physics, Southern University of Science and Technology, Shenzhen, Guangdong, 518055, China}

\author{S.~Sanna}
 \affiliation{Dipartimento di Fisica e Astronomia "A. Righi", Universit\'a di Bologna, I-40127 Bologna, Italy}

\author{H.~Luetkens}
\affiliation{Laboratory for Muon Spin Spectroscopy, Paul Scherrer Institute, CH-5232 Villigen PSI, Switzerland}

\author{Z.~Salman}
\affiliation{Laboratory for Muon Spin Spectroscopy, Paul Scherrer Institute, CH-5232 Villigen PSI, Switzerland}

\author{P.~Bonf\`a}
\email{pietro.bonfa@unipr.it}
\affiliation{Dipartimento di Scienze Matematiche, Fisiche e Informatiche, Universit(\`a) di Parma, I-43124 Parma, Italy}

\author{Z.~Guguchia}
\email{zurab.guguchia@psi.ch} 
\affiliation{Laboratory for Muon Spin Spectroscopy, Paul Scherrer Institute, CH-5232 Villigen PSI, Switzerland}

%\pacs{74.20.Mn, 74.25.Ha, 74.70.Xa, 76.75.+i}

\begin{abstract}

\end{abstract}

\maketitle

\newpage

\textbf{The breaking of time-reversal symmetry (TRS) in the normal state of kagome \cite{Syozi,ZHou,JXYin2,LYe,GuguchiaCSS,Mazin} superconductors $A$V$_{3}$Sb$_{5}$ stands out as a significant feature \cite{BOrtiz2,BOrtiz3,QYin,Wilson2023,YJiang,GuguchiaMielke,GuguchiaRVS,GuguchiaSc166,GuguchiaNPJ,KhasanovCVS,TNeupert,MDenner,YZhong,MHChristensen2022,GWagner,Grandi,GuoMoll,SYang,FYu}. Yet the extent to which this effect can be tuned remains uncertain, a crucial aspect to grasp in light of the varying details of TRS breaking observed through different techniques. Here, we employ the unique low-energy muon spin rotation technique combined with local field numerical analysis to study the TRS breaking response as a function of depth from the surface in single crystals of RbV$_{3}$Sb$_{5}$ with charge order and Cs(V$_{0.86}$Ta$_{0.14}$)$_{3}$Sb$_{5}$ without charge order. In the bulk (i.e., ${\textgreater}$ 33 nm from the surface) of RbV$_{3}$Sb$_{5}$, we have detected a notable increase in the internal magnetic field width experienced by the muon ensemble. This increase occurs only within the charge ordered state. Intriguingly, the muon spin relaxation rate is significantly enhanced near the surface (i.e., ${\textless}$ 33 nm from the surface) of RbV$_{3}$Sb$_{5}$, and this effect commences at temperatures significantly higher than the onset of charge order. Conversely, in Cs(V$_{0.86}$Ta$_{0.14}$)$_{3}$Sb$_{5}$, we do not observe a similar enhancement in the internal field width, neither in the bulk nor near the surface. These observations indicate a strong connection between charge order and TRS breaking on one hand, and on the other hand, suggest that TRS breaking can occur prior to long-range charge order. This research offers compelling evidence for depth-dependent magnetism in $A$V$_{3}$Sb$_{5}$ superconductors in the presence of charge order. Such findings are likely to elucidate the intricate microscopic mechanisms that underpin the TRS breaking phenomena in these materials.}

%%%%%%%%%%%%%%%%%%%%%%%%%%%%%%%%%%%%%%%%%%%%%%%%%%%%
\begin{figure*}[t!]
\centering
\includegraphics[width=1.0\linewidth]{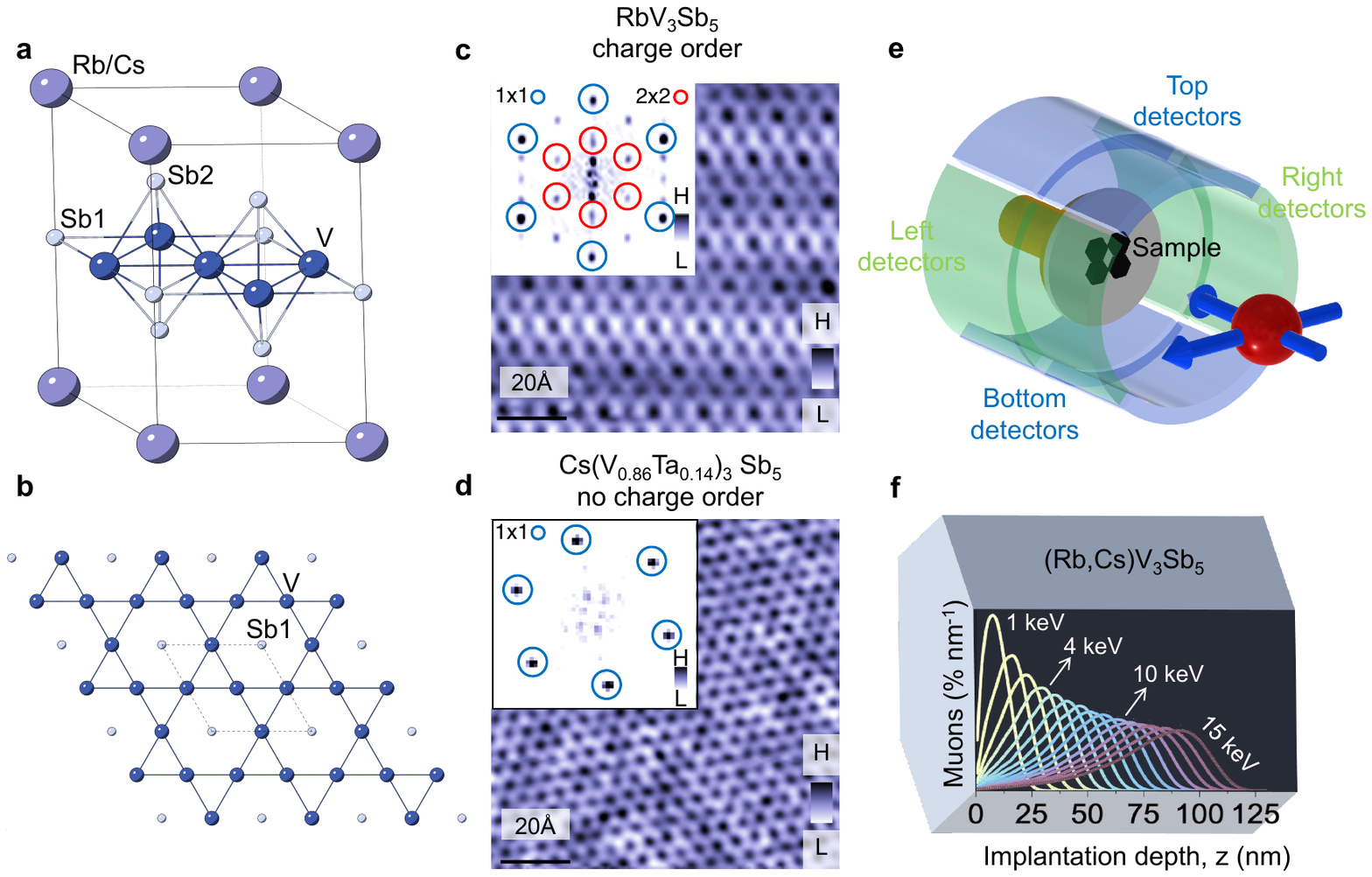}
\vspace{-1.0cm}
\caption{ (Color online) \textbf{Atomic topographic images of $A$V$_{3}$Sb$_{5}$ and experimental setup.} 
(a) The crystallographic structure of prototype compound $A$V$_{3}$Sb$_{5}$ ($A$ = Rb or Cs). The V atoms form a kagome lattice intertwined with a hexagonal lattice of Sb atoms. The (Rb,Cs) atoms occupy the interstitial sites between the two parallel kagome planes. In panel (b) the vanadium kagome net has been emphasized, with the interpenetrating antimony lattice included to highlight the unit cell (see dashed lines). (c) Scanning tunneling microscopy of the Sb surface for RbV$_{3}$Sb$_{5}$. The inset is the Fourier transform of this image, displaying 1${\times}$1 lattice Bragg peaks (blue circles) and 2${\times}$2 charge-order peaks (red circles). (d) Atomic topographic image of Sb surface in Ta-doped Cs(V$_{0.86}$Ta$_{0.14}$)$_{3}$Sb$_{5}$. The inset is the Fourier transform of this image, showing the absence of 2${\times}$2 ordering peaks, leaving only Bragg peaks. (e) Experimental LE-${\mu}$SR setup with applied field vector $B_{\rm ext}$ perpendicular to the sample surface (i.e., along the $c$-axis of RbV$_{3}$Sb$_{5}$), and arrays of positron detectors used to count muon decay events. (f) Muon implantation profile of (Rb,Cs)V$_{3}$Sb$_{5}$ simulated for several implantation energies.}
\label{fig2}
\end{figure*}
%%%%%%%%%%%%%%%%%%%%%%%%%%%%%%%%%%%%%%%%%%%%%%%%%%%%

The concept of chiral (TRS breaking) charge order is a fascinating aspect of modern condensed matter physics, reflecting a state where electron arrangements break mirror symmetry, akin to left-handed and right-handed twists. This chiral charge ordering can lead to unconventional electronic properties \cite{YJiang,GuoMoll}, and host exotic quasi-particles, potentially useful for quantum computing and novel electronic devices due to its influence on electronic band structures and interactions.

The $A$V$_{3}$Sb$_{5}$ kagome superconductors represent a unique and ideal platform known for hosting charge orders that possibly break TRS \cite{YJiang,GuguchiaMielke,GuguchiaRVS,GuguchiaNPJ,KhasanovCVS,LiYu,YXu,Balents,Nandkishore,Qimiao,CGuo}. 
This phenomenon of high-temperature TRS breaking charge order is exceedingly rare. It offers a profound analogy to pivotal theoretical models in physics: the Haldane model \cite{Haldane} for the honeycomb lattice and the Varma model \cite{Varma} for the square lattice. The evidence for TRS breaking in these materials primarily comes from zero-field and high-field muon spin rotation (${\mu}$SR) measurements. ${\mu}$SR, recognized for its exceptional sensitivity to magnetic phenomena \cite{GuguchiaNPJ,LukeTRS,HillierTRS,BernardoSRO}, has provided indications of TRS breaking in the charge ordered state of all three variants of $A$V$_{3}$Sb$_{5}$; KV$_{3}$Sb$_{5}$ \cite{GuguchiaMielke}, RbV$_{3}$Sb$_{5}$ \cite{GuguchiaRVS}, and CsV$_{3}$Sb$_{5}$ \cite{KhasanovCVS}. This uniformity across different compositions suggests that the nature of this symmetry breaking is intrinsic to Kagome superconductors in general.

In addition to  the increase of the internal magnetic fields in the TRS breaking state by ${\mu}$SR, other manifestations of unconventional charge order in these materials have been observed. Notably, these include the giant anomalous Hall effect \cite{SYang,FYu} detected through transport measurements.
Furthermore, a field-tunable chirality switch effect \cite{YJiang,NShumiya,Wang2021,YXing} and the ability to control chiral transport \cite{GuoMoll} properties have been reported. These features point to the presence of chiral electronic states, which are sensitive to external magnetic fields. Kerr effect measurements regarding TRS breaking are contradictory; some indicate its presence \cite{YXu}, while others do not \cite{Kapitulnik}. Among these various techniques, ${\mu}$SR stands out as arguably the most magnetically sensitive, providing a critical tool for detecting and understanding the subtle magnetic quantum phenomena occurring in these Kagome superconductors. The combination of high-temperature TRS breaking charge order and the unique set of properties in AV$_{3}$Sb$_{5}$ superconductors therefore not only challenge existing theoretical frameworks but also open up potential avenues for novel electronic applications \cite{Wilson2023}.

Previous ${\mu}$SR experiments \cite{GuguchiaMielke,GuguchiaRVS,GuguchiaNPJ,KhasanovCVS,LiYu} have primarily focused on exploring the TRS breaking response within the bulk of $A$V$_{3}$Sb$_{5}$ superconductors. However, there is a notable gap in our understanding regarding the magnetic characteristics near the surfaces of these materials. The  lack of understanding in this area is particularly crucial given the research on thin films \cite{SongB}, which shows non-monotonic variations in charge ordering temperature as a function of thickness. It is vital to comprehend the influence of the surface on the overall electronic and magnetic properties of the materials. Given this context, it becomes imperative to extend our investigations to probe the magnetism at the surface of $A$V$_{3}$Sb$_{5}$. Understanding surface magnetism is not only crucial for a comprehensive understanding of the material's magnetic properties, but also for unraveling the interplay between charge order and TRS breaking. Thus, a focused effort to explore and characterize the magnetic fingerprints at the surface of $A$V$_{3}$Sb$_{5}$ is essential to advance our knowledge in this area.

%%%%%%%%%%%%%%%%%%%%%%%%%%%%%%%%%%%%%%%%%%%%%%%%%%%%%%%%%%%%%%
\begin{figure*}[t!]
\centering
\includegraphics[width=1.0\linewidth]{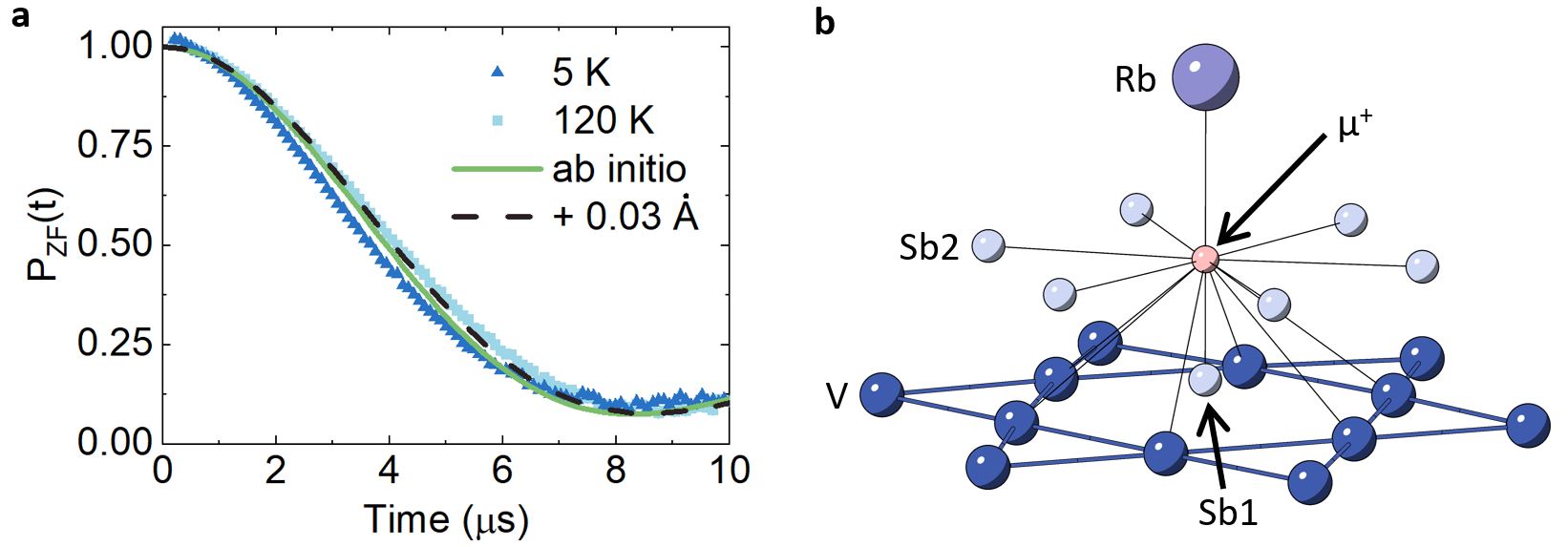}
\vspace{-0.5cm}
\caption{ (Color online) \textbf{Muon stopping site and ab initio prediction in RbV$_{3}$Sb$_{5}$.} 
(a) Zero-field $\mu$SR time spectra in the bulk of RbV$_3$Sb$_5$ at two temperatures, above and below $T_{\rm CO}$. The light green curve represents the ab initio prediction in the absence of electronic moments. The dashed black line is obtained by displacing the muon from its equilibrium position by only 0.03 ${\AA}$, which restores almost perfect agreement with the experimental results. (b) The muon site, indicated by the red ball located between the in-plane Sb1 and the Rb atom in RbV$_3$Sb$_5$, was obtained using the DFT+$\mu$ method. The figure also shows the displacement of the nearest neighbor Sb atom from the kagome plane formed by V atoms.} 
\label{fig2}
\end{figure*}
%%%%%%%%%%%%%%%%%%%%%%%%%%%%%%%%%%%%%%%%%%%%%%%%%%%%%%%%%%%%% 
 
%%%%%%%%%%%%%%%%%%%%%%%%%%%%%%%%%%%%%%%%%%%%%%%%%%%%%%%%%%%%%
\begin{figure*}[t!]
\centering
\includegraphics[width=1.0\linewidth]{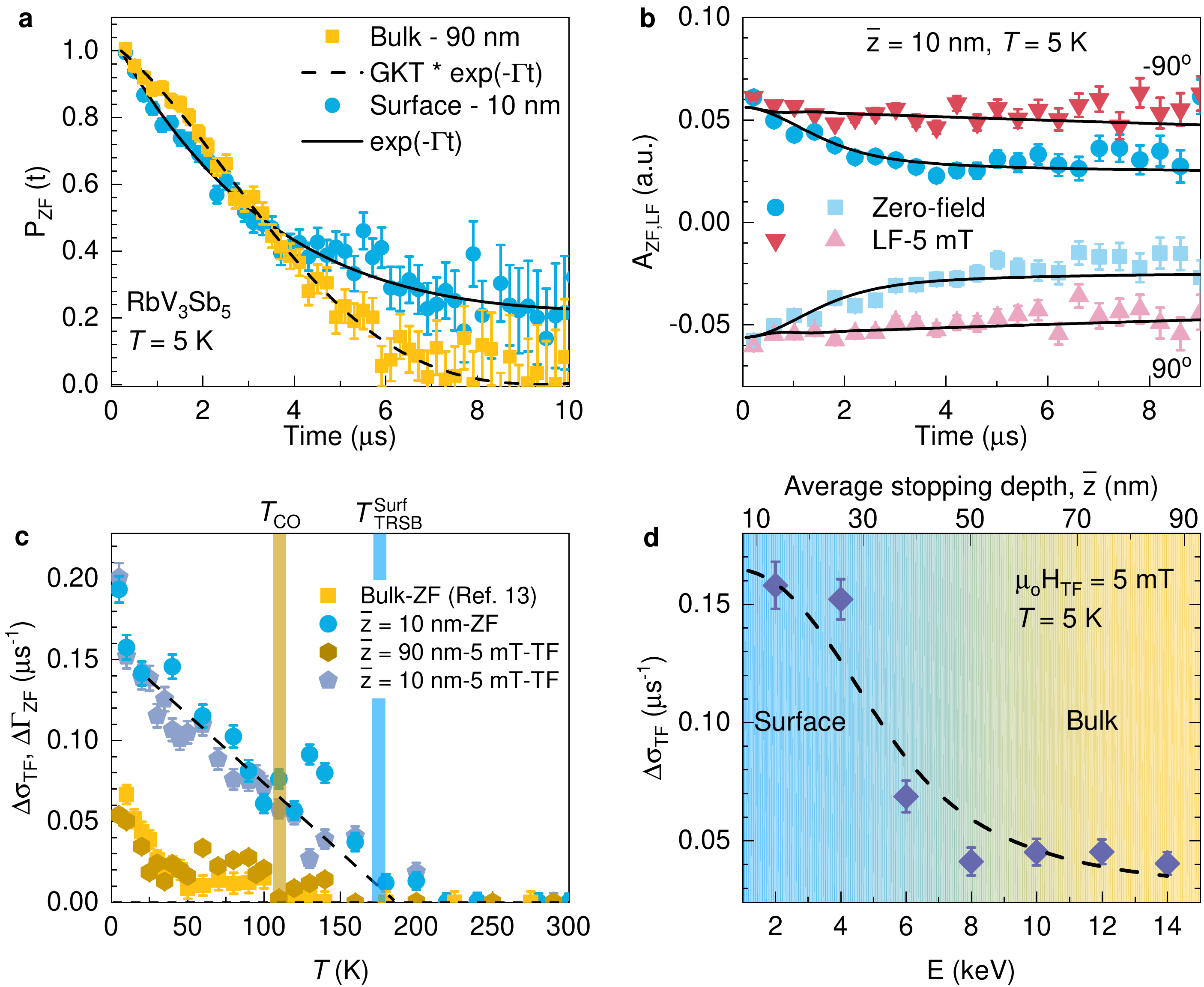}
\vspace{-0.5cm}
\caption{ (Color online) \textbf{Depth dependent magnetism in RbV$_{3}$Sb$_{5}$.} 
(a) The ZF ${\mu}$SR time spectra for the single crystal sample of RbV$_{3}$Sb$_{5}$, obtained at $T$ = 5 K at the surface (mean implantation depth of $\bar{z}$ = 10 nm) and in the bulk ($\bar{z}$ = 90 nm). The dashed curve represents a fit to Eq. (1). The solid line represents a fit using only the exponential function $\rm{exp}(-\Gamma t)$. (b) The ${\mu}$SR time spectra for RbV$_{3}$Sb$_{5}$, obtained at $T$ = 5 K in zero-field and under the magnetic field of 5mT for parallel (angle = $90^{\circ}$) and antiparallel (angle = -$90^{\circ}$) alignments of the applied field and the initial muon spin polarization. In both (a) and (c) the error bars are the standard error of the mean (s.e.m.) in about 10$^{6}$ events. (c) Temperature dependence of the relaxation rates ${\Delta}$${\Gamma}_{\rm ZF}$ = ${\Gamma}_{\rm ZF}(T)$ - ${\Gamma}_{\rm ZF}$(T=300K) and ${\Delta}$${\sigma}_{\rm TF}$ = ${\sigma}_{\rm TF}(T)$-${\sigma}_{\rm TF}$(300K), measured in zero-field and an applied field of 5 mT, respectively, at the surface ($\bar{z}$ ${\simeq}$ 10 nm) and in the bulk ($\bar{z}$ ${\simeq}$ 90 nm) of the single crystal RbV$_{3}$Sb$_{5}$. The absolute value of zero-field relaxation rate at room temperature is as follows: ${\sigma}_{ZF}$ $\simeq$ 0.153(5) ${\mu}s^{-1}$. (d) The muon-spin relaxation rate in RbV$_{3}$Sb$_{5}$, measured at 5 K and in applied field of 5 mT, as a function of muon implantation energy, $E$. Top axis shows the average implantation depth, $\bar{z}$. The dashed curve is the predicted behavior of the ${\Delta}$${\sigma}_{\rm TF}$ assuming a step-like depth dependence and considering the muon implantation profile \cite{Martins2023,Simoes}.}
\label{fig2}
\end{figure*}
%%%%%%%%%%%%%%%%%%%%%%%%%%%%%%%%%%%%%%%%%%%%%%%%%%%%%%%%%%%%%

In this study, we utilize the unique low-energy muon spin rotation method \cite{Morenzoni2002,Prokscha2008}, coupled with local field numerical analysis, to investigate the depth-dependent TRS breaking response in single crystals of RbV$_{3}$Sb$_{5}$ (which exhibits charge order) and Cs(V$_{0.86}$Ta$_{0.14}$)$_{3}$Sb$_{5}$ (which does not exhibit charge order). In RbV$_{3}$Sb$_{5}$, we detect a notable fourfold enhancement of the zero-field muon spin relaxation rate near the crystal's surface compared to the bulk. Calculations indicate that the observed increase in the relaxation rate is attributable to magnetism, rather than being a consequence of muon-induced structural distortions or a secondary effect due to structural changes stemming from charge order \cite{Bonfa}. In Cs(V$_{0.86}$Ta$_{0.14}$)$_{3}$Sb$_{5}$, which lacks charge order, there is no noticeable increase in the internal field width, both in the bulk and near the surface. These observations imply a strong connection between the TRS breaking response and the presence of charge order. This finding emphasizes the need for a microscopic understanding of why the surface offers more favorable conditions for the formation of novel magnetism.

The $A$V$_{3}$Sb$_{5}$ structure comprises a Kagome lattice of V atoms interlaced with a hexagonal lattice of Sb atoms, crystallizing in the $P$6/$mmm$ space group (Figs. 1a and b). Scanning tunneling microscopy (STM) atomic topographic images of the Sb surface for RbV$_{3}$Sb$_{5}$ and Cs(V$_{0.86}$Ta$_{0.14}$)$_{3}$Sb$_{5}$ single crystals are presented in Figures 1c and 1d, respectively. The Fourier transform of the image for RbV$_{3}$Sb$_{5}$ (inset of Fig. 1c) reveals both 1${\times}$1 lattice Bragg peaks (blue circles) and 2${\times}$2 charge-order peaks (red circles). In contrast, the Fourier transform for Cs(V$_{0.86}$Ta$_{0.14}$)$_{3}$Sb$_{5}$ displays only the Bragg peaks, indicating the absence of 2${\times}$2 ordering and thus charge order in this compound. Extended data Figure 1 shows that the critical temperature for superconductivity and the superfluid density are significantly higher in Cs(V$_{0.86}$Ta$_{0.14}$)$_{3}$Sb$_{5}$ compared to RbV$_{3}$Sb$_{5}$. This enhancement is attributed to the complete suppression of charge order in the Ta-doped sample. Figure 1e illustrates a schematic of the zero-field (ZF) and transverse-field low-energy ${\mu}$SR setup. 
Various detectors placed around the sample track the incoming $\mu^{+}$ and the outgoing $e^{+}$. We employed a muon beam with an adjustable energy range from $E$ = 1 keV to 30 keV. Each $E$ corresponds to a different muon implantation depth profile. This range of energies allows us to vary the implantation depths of the muons from a fraction of a nanometer up to 200 nm, therefore enabling us to conduct depth-dependent ${\mu}$SR studies (approximately $\bar{z}$ = 10 - 200 nm in depth). Figure 1f shows the muon implantation profile in (Rb,Cs)V$_{3}$Sb$_{5}$ for various implantation energies, simulated using the Monte Carlo algorithm TrimSP \cite{Morenzoni2002}.\\

%%%%%%%%%%%%%%%%%%%%%%%%%%%%%%%%%%%%%%%%%%%%%%%%%%%%%%%%%%%%%%%
\begin{figure*}[t!]
\centering
\includegraphics[width=1.0\linewidth]{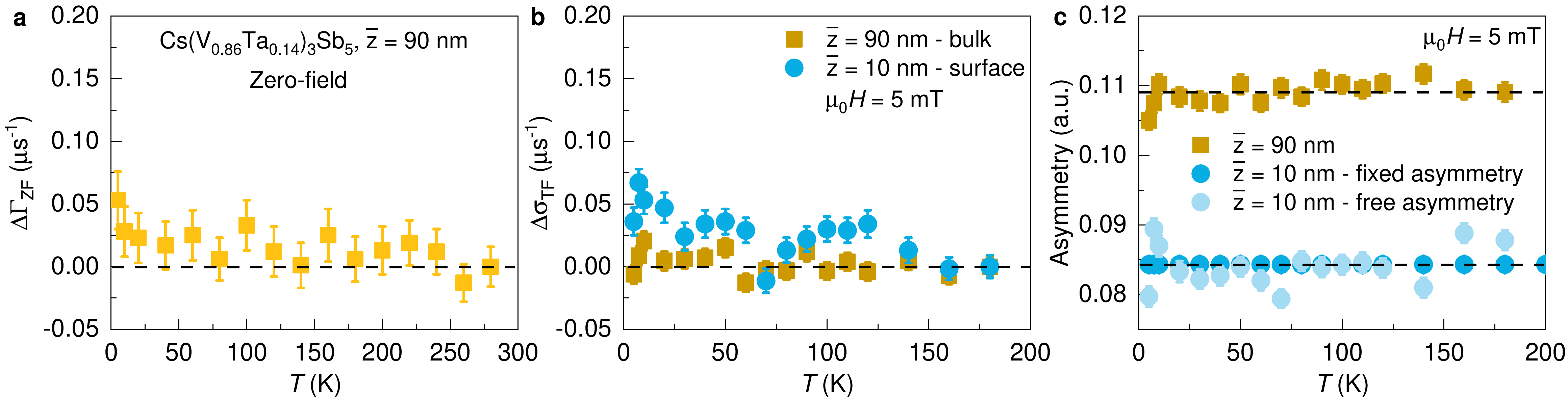}
\vspace{-0.5cm}
\caption{ (Color online) \textbf{Absence of magnetism in Cs(V$_{0.86}$Ta$_{0.14}$)$_{3}$Sb$_{5}$.} 
(a) Temperature dependence of the relaxation rate ${\Delta}$${\Gamma}_{\rm ZF}$ = ${\Gamma}_{\rm ZF}(T)$ - ${\Gamma}_{\rm ZF}$(T=300K), measured in the bulk ($\bar{z}$ ${\simeq}$ 90 nm) of the single crystal Cs(V$_{0.86}$Ta$_{0.14}$)$_{3}$Sb$_{5}$. (b-c) The temperature dependence of the low transverse field (5mT) muon spin relaxation rate ${\Delta}$${\sigma}_{\rm TF}$ = ${\sigma}_{\rm TF}(T)$-${\sigma}_{\rm TF}$(200K) (b) and the initial asymmetry (c), measured at the surface ($\bar{z}$ ${\simeq}$ 10 nm) and in the bulk ($\bar{z}$ ${\simeq}$ 90 nm) of the single crystal Cs(V$_{0.86}$Ta$_{0.14}$)$_{3}$Sb$_{5}$.} 
\label{fig2}
\end{figure*}
%%%%%%%%%%%%%%%%%%%%%%%%%%%%%%%%%%%%%%%%%%%%%%%%%%%%%%%%%%%%%%%

In Fig. 2a, we present high-statistics ZF-${\mu}$SR spectra for RbV$_{3}$Sb$_{5}$, displayed as the polarization function $P_{\rm ZF}(t)$ = $A_{\rm ZF}(t)$/$A_{\rm ZF}(0)$. These measurements were taken both above and below the charge ordering temperature $T_{\rm CO}$ ${\simeq}$ 110 K in the bulk of the material. A sizeable increase in the relaxation of the asymmetry observed at $T$ = 5 K is accompanied by a gradual change from a Gaussian-like curve to a more exponential-like curve for $P(t)$ at early times. The ZF-${\mu}$SR spectrum is well described using the Gaussian Kubo-Toyabe (GKT) depolarization function  \cite{Toyabe} multiplied by an exponential decay function, consistent with previous work \cite{GuguchiaMielke}:
\begin{equation}
\begin{aligned}
P_{ZF}^{\rm{GKT}}(t) =  \left(\frac{1}{3} + \frac{2}{3}(1 - \Delta_{ZF}^2t^2 ) \exp\Big[-\frac{\Delta_{ZF}^2t^2}{2}\Big]\right) \exp(-\Gamma_{ZF} t) \\ 
%\label{eq3}
\end{aligned}
\end{equation}  
where ${\Delta}$/${\gamma_{\mu}}$ represents the width of the local field distribution, arising from nuclear moments and ${\gamma_{\mu}}$/2${\pi}$ = 135.5~MHz/T is the muon gyromagnetic ratio. ${\Gamma}_{ZF}$ is attributed to muon spin relaxation originating from electronic sources \cite{GuguchiaMielke,GuguchiaNPJ}.

To obtain a quantitative understanding of the zero-field muon spin relaxation in RbV$_3$Sb$_5$, we utilized first-principles calculations to compute the muon polarization function, using Density Functional Theory (DFT) simulations within the DFT+$\mu$ approach \cite{BlundellSJ}. The first principles prediction of $P(t)$ accounts for the interaction between the muon and the  surrounding nuclear dipole moments. Additionally, the muon's impact on both the lattice and the electric field gradient at various nuclear sites was assessed. This is crucial since all isotopes in RbV$_3$Sb$_5$ undergo quadrupolar interactions. A comprehensive explanation of our methodology is available in the Supplementary Information. Our analysis reveals a single stable muon site in RbV$_3$Sb$_5$, depicted in Fig. 2b. The muon's nearest neighbor is the in-plane Sb1 atom, followed by an Rb atom, with two hexagonal arrangements of out-of-plane Sb2 and V atoms from the kagome lattice as further neighbors. We then calculated the muon polarization function, $P(t)$, using Celio's approach \cite{MCelio}, as implemented in the UNDI code \cite{PBonfa,Footnote}. The calculated polarization closely matches experimental observations, albeit it predicts a slightly faster depolarization. A perfect alignment with experimental data is achieved by minutely adjusting the muon's position, specifically a 0.03~\AA{} shift from the nearest Sb atom. Notably, similar discrepancies have been observed in other DFT+$\mu$ studies \cite{BlundellSJ,JMoller}. At $T$ $\sim$ $T_{\rm CO}$, the experimental and theoretical results align well, indicating that nuclear moments predominantly are responsible for the muon spin relaxation. However, the relaxation rates increase in the charge-ordered state indicating a deviation from static nuclear dipole-induced relaxation. Persisting with a nuclear-focused explanation for $P(t)$, the observed low-temperature variations imply significant shifts in the position or electronic environment of the muon's two nearest neighbors, Sb1 and Rb, within the charge-ordered state. This is not mirrored for Sb in CsV$_3$Sb$_5$, as evidenced by nuclear quadrupole resonance studies \cite{FengXY}, nor for Rb in RbV$_3$Sb$_5$ \cite{Frassineti}. This reinforces the conclusion \cite{GuguchiaMielke} that the observed increase in the relaxation rate at low temperatures is of electronic origin, and that the parameter ${\Gamma}_{ZF}$ in Eq. (1) reflects mostly the temperature-dependent electronic contribution. Therefore, an increase in ${\Gamma}_{ZF}$ signifies an increase of the local magnetic fields, i.e., the breaking of time-reversal symmetry.

We then examine the depth dependence of the time-reversal symmetry (TRS) breaking signal in both RbV$_{3}$Sb$_{5}$ and Cs(V$_{0.86}$Ta$_{0.14}$)$_{3}$Sb$_{5}$ crystals. This is achieved through ZF and low transverse field $\mu$SR experiments. For RbV$_{3}$Sb$_{5}$, the ZF-${\mu}$SR spectra measured at both the surface ($E$ = 2 keV, corresponding to a mean implantation depth, $\bar{z}$, of 10 nm) and in the bulk ($E$ = 14 keV, $\bar{z}$ = 90 nm) are shown in Fig. 3a. A noticeable difference in the shape of the field distribution was observed between these two depths. Specifically, the ZF-${\mu}$SR spectrum in the bulk was analyzed using the Eq. 1. Conversely, the ZF-${\mu}$SR spectrum at the surface is accurately described by only the exponential term. The zero-field relaxation, which is decoupled by applying a small external magnetic field longitudinally aligned with the muon spin polarization (B$_{LF}$ = 5 mT, where LF stands for ``Longitudinal Field``) (see Fig. 3b), suggests that the substantial relaxation observed at the surface is due to spontaneous fields that are static on the microsecond timescale \cite{Amatolecture,Dalmas}. The enhancement of the relaxation rate is also observed in weak transverse field (B$_{TF}$ = 5 mT) experiments (transverse-field ${\mu}$SR spectra are shown in Extended Data Figure 2). Figure 3c depicts the temperature dependence of the zero-field and transverse-field relaxation rates. This is represented in terms of the differences: ${\Delta}$${\Gamma}_{\rm ZF}$ = ${\Gamma}_{\rm ZF}(T)$ - ${\Gamma}_{\rm ZF}$(T=300K) and ${\Delta}$${\sigma}_{\rm TF}$ = ${\sigma}_{\rm TF}(T)$-${\sigma}_{\rm TF}$(300K). This approach is adopted to facilitate a clearer comparison between the two sets of data and to remove any potential systematic errors due to the different implantation energies used. The response observed at a depth of 90 nm from the surface is characterized by a two-step increase in the relaxation rate, commencing at the onset of the charge order temperature $T_{{\rm CO}}$ ${\simeq}$ 110 K. This finding aligns with previous results \cite{GuguchiaRVS} obtained using the GPS instrument \cite{AmatoGPS}, which predominantly probes the bulk response. Previously  \cite{GuguchiaRVS}, it was also shown that the two-step behavior becomes more pronounced when a high magnetic field is applied along the $c$-axis. At a shallower depth of 10 nm, the increase in the relaxation is about four times larger than in the bulk and decreases monotonously with increasing temperature, lacking the two-step feature. Interestingly, this rate tends to plateau at a temperature about 60 K higher than $T_{{\rm CO}}$ of the bulk. This observation suggests that not only does the magnitude of the magnetic response vary, but also the onset temperature shifts towards higher values at shallower depths. Specifically, the emergence of the TRS breaking signal near the surface occurs at $T^{\rm Surf}_{{\rm TRSB}}$ ${\simeq}$ 175 K. The relaxation rate as a function of implantation energy or mean depth $\bar{z}$, measured under a transverse field of 5 mT and at a temperature of 5 K, is depicted in Figure 3d. The energy dependence was fitted to the function \cite{Martins2023,Simoes}: 

\begin{equation}
{\Delta}{\sigma}_{TF}(E) = \int_0^\infty P(z,E){\Delta}\sigma(z) dz.
\end{equation}

where $P(z,E)$ is the probability of the muon beam implanted with energy $E$ to stop at a depth $z$, shown in Fig. 1f. 
${\Delta}$${\sigma}_{\rm TF}$ is assumed to have a step-like function with two regions. This analysis reveals a characteristic depth of $\bar{z}_{c}$ ${\simeq}$ 33 nm, in which we observe a notable enhancement in the relaxation rate. This finding is significant as it establishes a characteristic depth where the materials properties begin to exhibit marked changes, distinguishing the near surface behavior from that of the bulk.

We now turn our attention to the results for Cs(V$_{0.86}$Ta$_{0.14}$)$_{3}$Sb$_{5}$, which are summarized in Figure 4. 
Contrary to RbV$_{3}$Sb$_{5}$, the variation in the field distribution shape between these two depths is much less pronounced. Figure 4a displays the temperature dependence of the zero-field relaxation rates ${\Delta}_{\rm ZF}$ and ${\Gamma}_{\rm ZF}$, as measured in the bulk at a depth of 90 nm. In this case, both rates remain constant across the entire temperature range, with no discernible increase at lower temperatures. To minimize data scatter in ${\Gamma}_{\rm ZF}$, ${\Delta}_{\rm ZF}$ was held common to all temperatures in our final analysis. These results indicate that in Cs(V$_{0.86}$Ta$_{0.14}$)$_{3}$Sb$_{5}$, where charge order is fully suppressed, there is an absence of a TRS breaking response. Additionally, as depicted in Figure 4b, there is  no enhancement in the relaxation rate near the surface. This is evident when comparing the transverse field (TF) relaxation rates at depths of $\bar{z}$ = 90 nm and $\bar{z}$ = 10 nm. The initial asymmetry also exhibits no temperature dependence down to 5K (see Fig. 4c), further confirming the lack of any low-temperature anomalies in Cs(V$_{0.86}$Ta$_{0.14}$)$_{3}$Sb$_{5}$.

Previous ${\mu}$SR research \cite{GuguchiaMielke,GuguchiaRVS,KhasanovCVS} uncovered weak internal magnetic fields of around 0.6 G in the charge ordered state of $A$V$_{3}$Sb$_{5}$, hinting at spontaneous time-reversal symmetry breaking. However, the faintness of internal fields has often prompted inquiries into their intrinsic nature. In this paper, we report four key findings:
\textbf{(1)} Through muon stopping site calculations and local field numerical analysis, we have ruled out muon-induced effects or structural distortions as the cause for the increased zero-field muon spin relaxation rate, attributing it instead to intrinsic magnetism. The observation of TRS-breaking charge order in $A$V$_{3}$Sb$_{5}$ has been ascribed to orbital current order within the vanadium kagome layer \cite{MDenner,MHChristensen2022,Balents,Nandkishore}, potentially exerting a significant influence on the superconducting state. Theoretical models indicate an exceedingly small net flux, resulting in a correspondingly minor net magnetic moment within the unit cell of the orbital current order. The hypothesized orbital current is thought to be consistent throughout the lattice, but with an alternating flow direction, leading to non-uniform fields at the muon site. In this context, muons could interact with these closed current loops below the temperature $T^{\rm Surf}_{{\rm TRSB}}$, which would result in an enhanced internal field width as detected by the muon ensemble, in the charge ordered state. Similar effects of static magnetic fields appearing near the surface due to orbital loop currents were observed in Sr$_{2}$RuO$_{4}$ crystals \cite{BernardoSRO}. The muon stopping site is notably distant (${\sim}$ 3.5~\AA{}) from the vanadium lattice, symmetrically situated around the hexagon of vanadium atoms. This symmetric arrangement, together with the negligible net flux resulting from the orbital currents,  can account for the observed small TRS breaking signal in the charge-ordered state.
\textbf{(2)} A pronounced enhancement, by a factor of four, of the zero/low-field muon spin relaxation rate is observed near the surface of
RbV$_{3}$Sb$_{5}$ compared to the bulk. The characteristic depth scale at which the enhancement of relaxation occurs is $\bar{d}_{c}$ 
${\simeq}$ 33 nm. Near the surface, i.e. below 40 nm, the estimated field strength is ${\Gamma}$$_{12}$/${\gamma_{\mu}}$ ${\simeq}$ 2.5~G, whereas in the bulk it is 0.6 G. This not only provides stronger evidence of TRS breaking in this material but also demonstrates the significant tunability of the TRS signal under zero-field conditions. \textbf{(3)} Near the surface, the onset of the TRS breaking response seems to occur at a temperature $T^{\rm Surf}_{{\rm TRSB}}$ ${\simeq}$ 175 K. This indicates that in the bulk of RbV$_{3}$Sb$_{5}$, TRS breaking takes place within the charge ordered state, whereas near the surface, it emerges at a temperature notably higher than the onset of charge order $T_{{\rm CO}}$ ${\simeq}$ 110 K. Given that a range of surface and bulk-sensitive methods identify 110 K as the onset temperature for charge order in RbV$_{3}$Sb$_{5}$, it can be logically inferred that the temperature at which charge order occurs is consistent across both the surface and the bulk. \textbf{(4)} Furthermore, in Cs(V$_{0.86}$Ta$_{0.14}$)$_{3}$Sb$_{5}$, which lacks charge order, no increase in the relaxation rate is observed either at the surface or in the bulk down to 5 K. This strongly suggests a direct correlation between charge order and the TRS breaking signal in $A$V$_{3}$Sb$_{5}$ Kagome superconductors and rules out systematic effects as a source of the relaxation enhancement near the surface. These findings also show that while the TRS breaking response is closely related to the presence of charge order, TRS breaking can manifest at temperatures higher than those of charge order onset. In this regard, recent torque measurements \cite{TAsaba} have demonstrated a two-fold in-plane magnetic anisotropy above charge order , which breaks the rotational symmetry of the crystal. This finding aligns with our observations. Such insights not only broaden our understanding of the distinct quantum states in these superconductors but also open avenues for further exploration into the realm of unconventional superconductivity.

Our research identifies a kagome superconductor RbV$_{3}$Sb$_{5}$ as the system with the highest TRS breaking temperature, reaching ${\simeq}$ 175 K. The observation that the TRS breaking signal at the surface of RbV$_{3}$Sb$_{5}$ occurs at a higher temperature than the onset of charge order presents an intriguing and novel aspect of the physics in these materials. This suggests that the mechanism driving the TRS breaking phenomenon might be different or more pronounced near the surface. This could mean that surface interactions or reconstructions play a significant role, possibly indicating an enhanced or modified electron correlation effect near the surface compared to the bulk. Typically, it is anticipated that surface effects occur extremely close to the surface. Present investigations suggest that the transition from bulk to surface occurs over a range of 33 nm in RbV$_{3}$Sb$_{5}$. This observation might open avenues for tuning the electronic properties of these materials through surface engineering, which could be relevant for potential applications in electronic devices where surface properties are crucial. This finding also cautions that surface-sensitive techniques such as ARPES or STM may yield insights that differ from those obtained using bulk-sensitive methods. The fact that TRS breaking occurs at a higher temperature than charge order in RbV$_{3}$Sb$_{5}$ mirrors the behavior observed in cuprate high-temperature superconductors \cite{Keimer}, where the pseudogap phase, thought to involve orbital current \cite{Varma}, also emerges at a higher temperature than charge order. This similarity draws intriguing parallels between these two distinct types of materials and points to potentially fundamental and universal behaviors in these complex material systems.

\section{METHODS}

\textbf{Muon-Spin Rotation}: In a ${\mu}$SR (muon spin rotation) experiment, nearly 100${\%}$ spin-polarized muons ${\mu}^{+}$ are implanted into the sample one at a time. These positively charged ${\mu}^{+}$ particles thermally stabilize at interstitial lattice sites, effectively serving as magnetic microprobes within the material. In the presence of a magnetic field, the muon spin undergoes precession at the local field $B_{\rm \mu}$ at the muon site, with a Larmor frequency ${\nu}_{\rm \mu}$ given by $\gamma_{\rm \mu}$/(2${\pi})$$B_{\rm \mu}$, where $\gamma_{\rm \mu}$/(2${\pi}$) = 135.5 MHz T$^{-1}$ represents the muon gyromagnetic ratio.\\

\textbf{Experimental details}: Zero field (ZF) and weak transverse field (TF) $\mu$SR experiments were conducted on single crystalline samples of RbV$_{3}$Sb$_{5}$ and Cs(V$_{0.86}$Ta$_{0.14}$)$_{3}$Sb$_{5}$ using the low energy $\mu$SR instrument at the Swiss Muon Source (S$\mu$S), Paul Scherrer Institut, in Villigen, Switzerland \cite{Morenzoni2002,Prokscha2008}. For these measurements, large single crystal pieces were used. The crystals were carefully arranged in a mosaic layout on a nickel-coated plate and secured with silver epoxy, covering an area of 1.5 $\times$ 1.5 cm$^{2}$. The samples were mounted on a cold finger cryostat, which accommodates temperatures ranging from 5-300 K. The crystals were aligned such that was done such that the $c$-axis was parallel to the muon beam and the applied magnetic field. Measurements were carried out with the muon spin polarization both parallel to the $c$-axis (in a longitudinal configuration) and perpendicular to the $c$-axis (in a transverse field configuration). We utilized a muon beam that could be adjusted within an energy range of 1 keV to 30 keV. The implantation energy, $E$, corresponds to a specific muon implantation depth profile, allowing us to vary the implantation depths from a few nanometers to several tens nanometers. This capability facilitated our depth-resolved $\mu$SR studies, with approximate depths ranging from $\bar{z}$ = 1-200 nm. The muon implantation profiles in (Rb,Cs)V$_{3}$Sb$_{5}$ for various implantation energies, were simulated using the TrimSP Monte Carlo algorithm \cite{Morenzoni2002}.\\

Measurements of both normal and superconducting bulk state properties were conducted using the GPS and the high-field HAL-9500 instruments. HAL-9500 is equipped with a BlueFors vacuum-loaded, cryogen-free dilution refrigerator (DR) for probing the low ttemperature deep bulk properties. At the GPS instrument (${\pi}$M3) beamline, we utilized a "spin rotator" to alter the muon's spin orientation. Typically, a muons spin is naturally antiparallel to its momentum, but we rotated it by $45^{\circ}$ relative to the $c$-axis of the crystal. This allowed the samples orientation to remain fixed while the muon spin was adjusted. The rotation angle of $44.5(3)^{\circ}$ was precisely determined through measurements in a weak magnetic field, applied transversely to the muon spin polarization.\\

The ${\mu}$SR data, shown in Fig. 2a, were taken at the ISIS Pulsed Neutron and Muon Source at the Rutherford Appleton Laboratories (UK) using the EMU spectrometer. The powder sample was pressed into a disk of 30 mm in diameter and 1.9 mm in thickness. A 1.25 mm thick Kapton mask was mounted on top of the aluminum sample holder in order to eliminate the background signal originating from muons implanted in the sample holder. It is reasonable to assume that a 1${\%}$ background is still present in the measurements but, for the sake of simplicity, it is not removed from the experimental asymmetry. Measurements were carried out at 120~K, above the charge ordering transition, and at 5~K, well below $T_{CO}$. The total asymmetry is determined through transverse field calibration measurements at each temperature, and it is subsequently employed to assess the zero-field total asymmetry. Alternatively, this information can be extracted by fitting the asymmetry with a Kubo-Toyabe function in the interval 0.4 ${\mu}$s to 4${\mu}$s. The two approaches yield very similar results differing by less than 1${\%}$. This procedure allows for the experimental determination of the muon's spin polarization as a function of time.\\

\textbf{Analysis of transverse field $\mu$SR data}:
Muon spin rotation data were processed utilizing the software Musrfit, which was developed at the Paul Scherrer Institute \cite{Bastian}. The  TF-${\mu}$SR data were analyzed by using the following functional form \cite{Bastian}:

%%%%%%%%%%%%%%%%%%%%%%%%%%%%%%%%%%%%%%%%%%%%%%%%%%%%%%%%%%%%%%%
\begin{equation}
A_{TF}(t)=A_Se^{\Big[-\frac{\sigma_{TF}^2t^2}{2}\Big]}\cos(\gamma_{\mu}B_{\rm int}t+\varphi), 
\label{eq1}
\end{equation}
Here $A_{\rm S}$ denotes the initial assymmetry, and ${\varphi}$ is the initial phase of the muon-spin ensemble.
$B_{\rm int}$ represents the internal magnetic field at the muon site, and the relaxation rate ${\sigma}_{\rm TF}$ 
characterize the damping of the ${\mu}$SR signal.\\ 

\textbf{Computational details}: The muon sites in RbV$_{3}$Sb$_{5}$ have been investigated with Density Functional Theory (DFT) using the
so called DFT+$\mu$ \cite{BlundellSJ} approach.
The tri-hexagonal lattice structure \cite{MiaoPRB,HTan} was used to perform all simulations unless otherwise specified. This structure has $Fmmm$ symmetry with lattice parameters set to 10.943, 18.954 and 18.146~\AA{} for $a,b$ and $c$, respectively \cite{OrtizPRM}.
The simulations were carried out using the plane wave based code QuantumESPRESSO v7.1~\cite{Giannozzi}
The structural relaxation of all lattice structures was performed with GBRV ultrasoft pseudopotentials~\cite{GarrityCMS}
using 40~Ry cutoff for the planewave expansion of wavefunctions and 320~Ry cutoff
for the charge density. The PBEsol\cite{PerdewPRL} functional was used to estimate the
exchange and correlation term. The reciprocal space was sampled with the Gamma point.
The optimization of atomic coordinates was carried out until forces and total energy differences were less than
0.5~mRy/Bohr and 0.09~mRy respectively. In order to find the stable muon sites we sampled the interstitial space of the host lattice using a grid with 1.2 \AA{} spacing between the points and removed all symmetry equivalent positions as well as all
points closer than 1.3 \AA{} to the atoms of RbV$_{3}$Sb$_{5}$. This results in 71 starting interstitial positions. After structural relaxations, 34 symmetrically inequivalent positions are found using the clustering algorithm available in Ref.~\cite{Onuorah2023}. For the stable muon sites a refined equilibrium position and the Electric Field Gradients (EFG) at the nuclei of the lattice, obtained with the GIPAW code~\cite{Ceresoli}, are estimated with a $2 \times 1 \times 1$ supercell.\\

%%%%%%%%%%%%%%%%%%%%%%%%%%%%%%%%%%%%%%%%%%%%%%%%%%%%%%%%%%%%%%%%%%%%%%
\section{Acknowledgments}~
The ${\mu}$SR experiments were carried out at the Swiss Muon Source (S${\mu}$S) Paul Scherrer Insitute, Villigen, Switzerland. 
Some of the data were also collected at the ISIS facility, STFC Rutherford Appleton Laboratory, UK. Authors acknowledge Peter Baker and Rhea Stewart for technical support and fruitful discussions. Z.G. acknowledges support from the Swiss National Science Foundation (SNSF) through SNSF Starting Grant (No. TMSGI2${\_}$211750). S.D.W. and A.C.S.  gratefully acknowledge support via the UC Santa Barbara NSF Quantum Foundry funded via the Q-AMASE-i program under award DMR-1906325.\\

%\textbf{Author Contributions}
%Z. Guguchia conceived and supervised the project. Growth of the single crystals RbV$_{3}$Sb$_{5}$ and Cs(V$_{0.86}$Ta$_{0.14}$)$_{3}$Sb$_{5}$: Y.Z., K.O., S.W., A.C.S. AND Z.W.. Low energy muon spin rotation experiments, analysis and corresponding discussions: J.N.G, C.M.III, D.D., M.M., V.S., M.F., T.T., J.-X.Y., M.Z.H., H.L., R.K., S.S., P.B., A.S., T.P., Z.S., and Z.G.. Muon stopping cite calculations, local field numerical analysis and corresponding discussions: T.M., S.S., Z.G., and P.B.. Figure development and writing of the paper: Z.G. with contributions from J.N.G, P.B. and other authors. All authors discussed the results, interpretation and conclusion.\\

%\includepdf[pages={-}]{MoTe2Supplementary.pdf}
%\includepdf[pages={3}]{MoTe2Supplementary.pdf}
%\includepdf[pages={4}]{MoTe2Supplementary.pdf}

\newpage
\newpage

\newpage

\renewcommand{\figurename}{Supplementary Figure}
\section{SUPPLEMENTAL MATERIAL \\ }

\subsection{Distinct superfluid response in RbV$_{3}$Sb$_{5}$ and Cs(V$_{0.86}$Ta$_{0.14}$)$_{3}$Sb$_{5}$}

Superfluid density of RbV$_{3}$Sb$_{5}$ and Cs(V$_{0.86}$Ta$_{0.14}$)$_{3}$Sb$_{5}$ was measured using Transverse field (TF) $\mu$SR experiments. TF-$\mu$SR experiments on the single crystalline samples of RbV$_{3}$Sb$_{5}$ were performed on the high-field HAL-9500 instrument, equipped with BlueFors vacuum-loaded cryogen-free dilution refrigerator (DR), at the Swiss Muon Source (S$\mu$S) at the Paul Scherrer Institut, in Villigen, Switzerland. TF-$\mu$SR experiments on the single crystalline samples of Cs(V$_{0.86}$Ta$_{0.14}$)$_{3}$Sb$_{5}$ were performed on the Dolly instrument. The $^{4}$He cryostats equipped with the $^{3}$He insets (base temperature $\simeq0.25$~K) were used. 

Supplementary Figure 5 displays the temperature dependence of the inverse squared penetration depth ${\lambda}_{ab}^{-2}$ for single crystals of RbV$_{3}$Sb$_{5}$ with charge order and Cs(V$_{0.86}$Ta$_{0.14}$)$_{3}$Sb$_{5}$ without charge order. These measurements were conducted in applied magnetic fields of  ${\mu}_{\rm 0}H = 5$~mT and ${\mu}_{\rm 0}H = 10$~mT, respectively. The onset of the superconducting transition was estimated to be approximately 0.8 K for RbV$_{3}$Sb$_{5}$ and about 4.7 K for Cs(V$_{0.86}$Ta$_{0.14}$)$_{3}$Sb$_{5}$. Notably, both the superconducting transition temperature and the superfluid density are significantly higher in Cs(V$_{0.86}$Ta$_{0.14}$)$_{3}$Sb$_{5}$ compared to RbV$_{3}$Sb$_{5}$. This enhancement is attributed to the complete suppression of charge order in the former compound. Crucially, there is a correlation between $T_{{\rm c}}$ and superfluid density, which is a characteristic feature of unconventional superconductivity. The temperature dependence of ${\lambda}_{ab}^{-2}(T)$ for RbV$_{3}$Sb$_{5}$ is consistent with a model featuring a nodal superconducting gap, whereas for Cs(V$_{0.86}$Ta$_{0.14}$)$_{3}$Sb$_{5}$, it aligns with a nodeless superconducting gap. This observation is in agreement with our previous findings that the suppression of charge order leads to a distinctly nodeless behavior in superfluid density.

%%%%%%%%%%%%%%%%%%%%%%%%%%%%%%%%%%%%%%%%%%%%%%%%%%%%%%%%%%%%%
\begin{figure}[t!]
\centering
\includegraphics[width=1.0\linewidth]{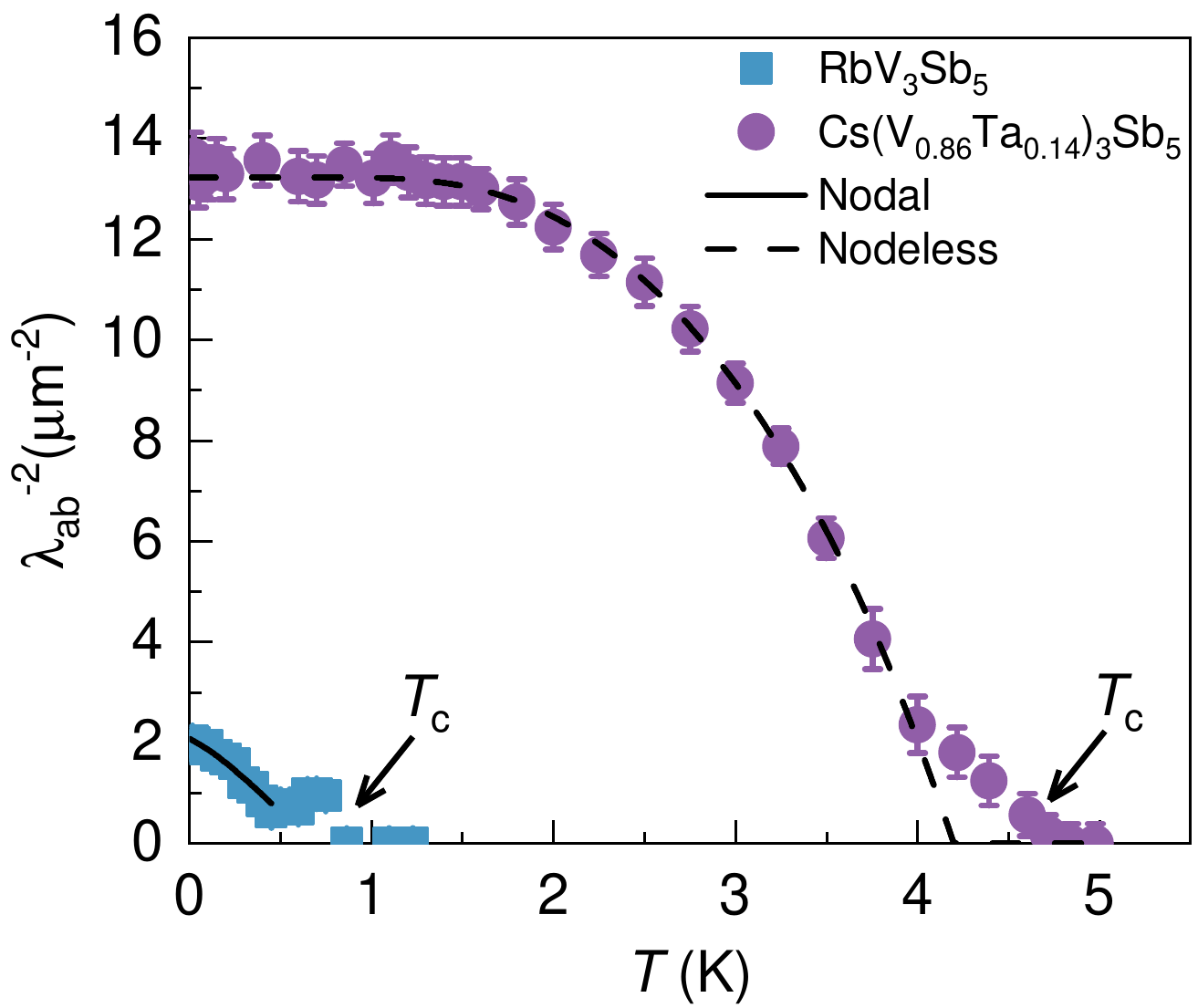}
\vspace{-0.5cm}
\caption{ (Color online) \textbf{Superfluid density.} 
The temperature dependence of the  inverse squared penetration depth ${\lambda}_{ab}^{-2}$ for the single crystals of RbV$_{3}$Sb$_{5}$ and Cs(V$_{0.86}$Ta$_{0.14}$)$_{3}$Sb$_{5}$, measured in an applied magnetic field of ${\mu}_{\rm 0}H = 5$~mT and ${\mu}_{\rm 0}H = 10$~mT, respectively. The error bars represent the standard deviations of the fit parameters. The solid (dashed) lines correspond to a fit using a model with nodal (nodeless) superconducting gap. Adapted from Ref. \cite{GuguchiaRVS2023}}
\label{fig2}
\end{figure}
%%%%%%%%%%%%%%%%%%%%%%%%%%%%%%%%%%%%%%%%%%%%%%%%%%%%%%%%%%%%

%%%%%%%%%%%%%%%%%%%%%%%%%%%%%%%%%%%%%%%%%%%%%%%%%%%%%%%%%%%%%
\begin{figure*}[t!]
\centering
\includegraphics[width=1.0\linewidth]{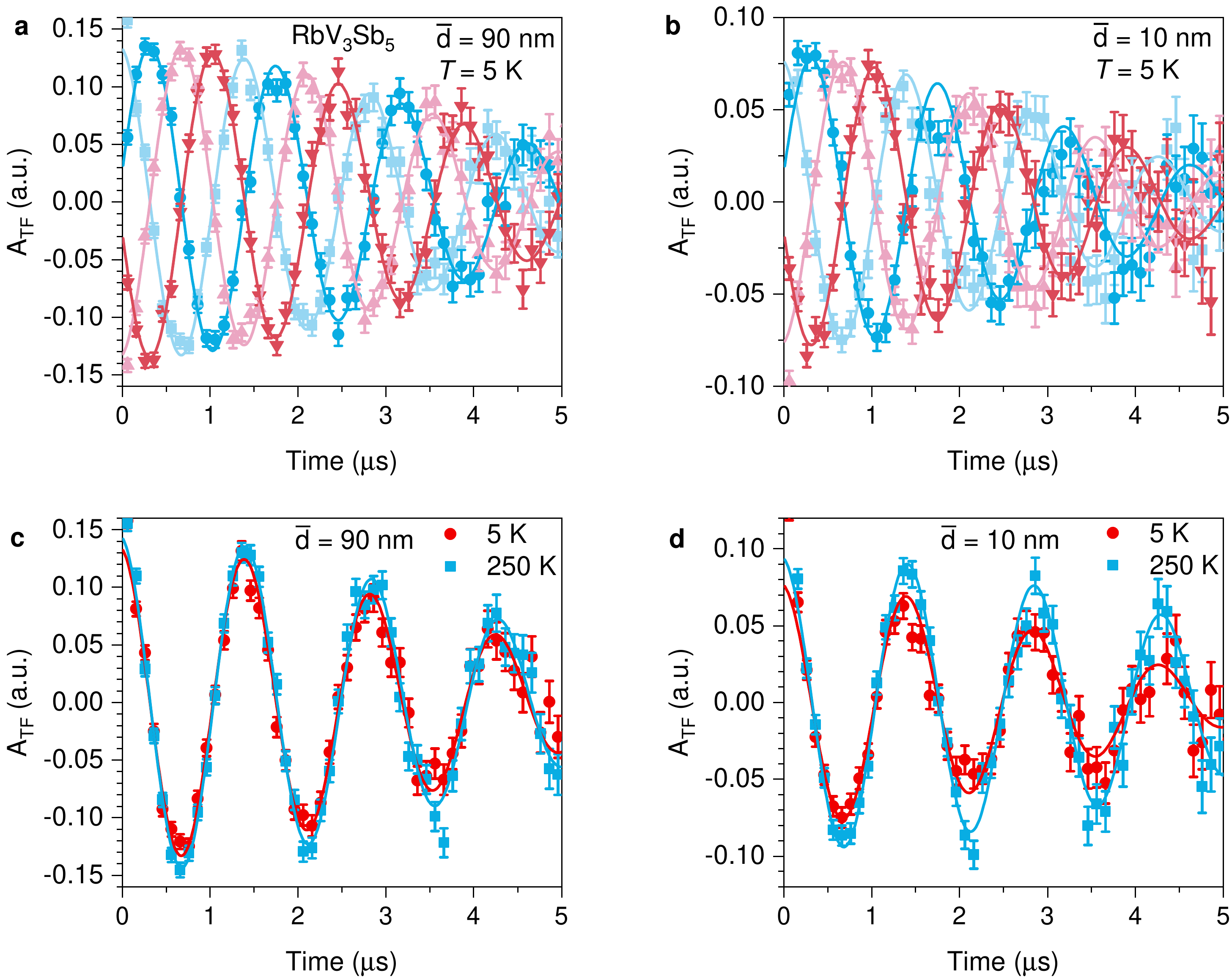}
\vspace{-0.5cm}
\caption{ (Color online) \textbf{Transverse-field (TF) ${\mu}$SR time spectra of RbV$_{3}$Sb$_{5}$.} 
(a-b) ${\mu}$SR spectra, measured near the surface at $\bar{d}$ = 10 nm and in the bulk at $\bar{d}$ = 90 nm of the single crystal RbV$_{3}$Sb$_{5}$ at $T$=5 K in a low field of ${\mu}_{\rm 0}H = 5$~mT, applied transversely to the muon spin polarization. (c) TF-${\mu}$SR spectra, measured in the bulk at both 250 K and 5 K. (d) TF-${\mu}$SR spectra, measured near the surface at 250 K and 5 K. Error bars represent the standard error of the mean (s.e.m.) in approximately 10$^{6}$ events. The error of each bin count  $n$ is determined by its standard deviation (s.d.). The errors for each bin in $A(t)$ are then calculated using standard error propagation.}
\label{fig2}
\end{figure*}
%%%%%%%%%%%%%%%%%%%%%%%%%%%%%%%%%%%%%%%%%%%%%%%%%%%%%%%%%%%%

\subsection{Transverse-field (TF) ${\mu}$SR time spectra of RbV$_{3}$Sb$_{5}$}

Supplementary Figure 6 presents the \(\mu\)SR spectra of the single crystal RbV\(_{3}\)Sb\(_{5}\), measured in the bulk at an average depth (\(\bar{d}\)) of 90 nm and near the surface at \(\bar{d}\) = 10 nm, respectively. These measurements were taken at a temperature of \(T\) = 5 K in a low magnetic field of \(\mu_0H = 5\) mT, applied transversely to the muon spin polarization. Extended Data Figure 2c displays a comparison of the TF-\(\mu\)SR spectra measured in the bulk at 5 K and 250 K, revealing a noticeable difference in the relaxation rate. This difference is even more pronounced at the surface, as illustrated in panel d.

\subsection{Muon stopping site and local field numerical analysis in \rvb}
\subsubsection{Computational details}

The muon sites in \rvb have been investigated with Density Functional Theory (DFT) using the
so called DFT+$\mu$ \cite{BlundellSJ} approach.
The tri-hexagonal lattice structure \cite{MiaoPRB,HTan} was used to perform all simulations.
This structure has $Fmmm$ symmetry with lattice parameters set to
10.943, 18.954 and 18.146~\AA{} for $a,b$ and $c$, respectively \cite{OrtizPRM}.
The simulations were carried out using the plane wave based code QuantumESPRESSO v7.1~\cite{Giannozzi}.
The structural relaxation of all lattice structures was performed with GBRV ultrasoft pseudopotentials~\cite{GarrityCMS}
using 40~Ry cutoff for the planewave expansion of wavefunctions and 320~Ry cutoff
for the charge density.
The PBEsol~\cite{PerdewPRL} functional was used to estimate the
exchange and correlation term. The reciprocal space was sampled with the Gamma point.
The optimization of atomic coordinates was carried out until forces and total energy differences were less than
0.5~mRy/Bohr and 0.09~mRy respectively.
For the bulk material, the atomic positions shown in Supplementary Table~\ref{tab:pos} closely resemble those reported in Ref.~\onlinecite{SubediPRM}.
In order to find the stable muon sites we sampled the interstitial space of the host lattice using a grid with 1.2 \AA{} spacing
between the points
and removed all symmetry equivalent positions as well as all
points closer than 1.3 \AA{} to the atoms of \rvb.
This results in 71 starting interstitial positions.
After structural relaxations, 34 symmetrically inequivalent positions are found using the clustering algorithm available in Ref.~\cite{Onuorah2023}.
They can be inspected in the online repository https://archive.materialscloud.org/record/2024.30.

For the stable muon sites a refined equilibrium position and the Electric
Field Gradients (EFG) at the nuclei of the lattice, obtained with the GIPAW code~\cite{Ceresoli}, are estimated with a
$2 \times 1 \times 1$ supercell.

\begin{table}[]
    \centering
    \begin{tabular}{SSSS} \toprule
\hline
    Atom & $x$ & $y$ & $z$ \\ \midrule
    \hline
        Rb & 0& 0& 0.253 \\
        Rb & 0& 0& 0.25 \\ \midrule
        \hline
        V  & 0.119& 0.377& 0 \\
        V  & 0.885& 0.127& 0 \\
        V  & 0.746& 0& 0 \\
        V  & 0& 0.246& 0 \\ \midrule
       \hline
       Sb  &  0&  0.830&  0.622  \\
       Sb  &  0.752&  0.084&  0.120  \\
       Sb  &  0&  0.331&  0.627  \\
       Sb  &  0&  0&   0.5   \\
       Sb  &  0.25&  0.25& 0     \\
       Sb  &  0&  0&   0     \\
       Rb  &  0&  0&   0.253 \\ \bottomrule
    \hline
    \end{tabular}
    \caption{Optimized positions in fractional coordinates for the unperturbed \rvb lattice with parameters $a=10.943$, $b=18.954$, $c=18.146$ and spacegroup Fmmm (no. 69).}
    \label{tab:pos}
\end{table}

\subsubsection{Muon sites}
\begin{figure*}
\centering
\includegraphics[width=2\columnwidth]{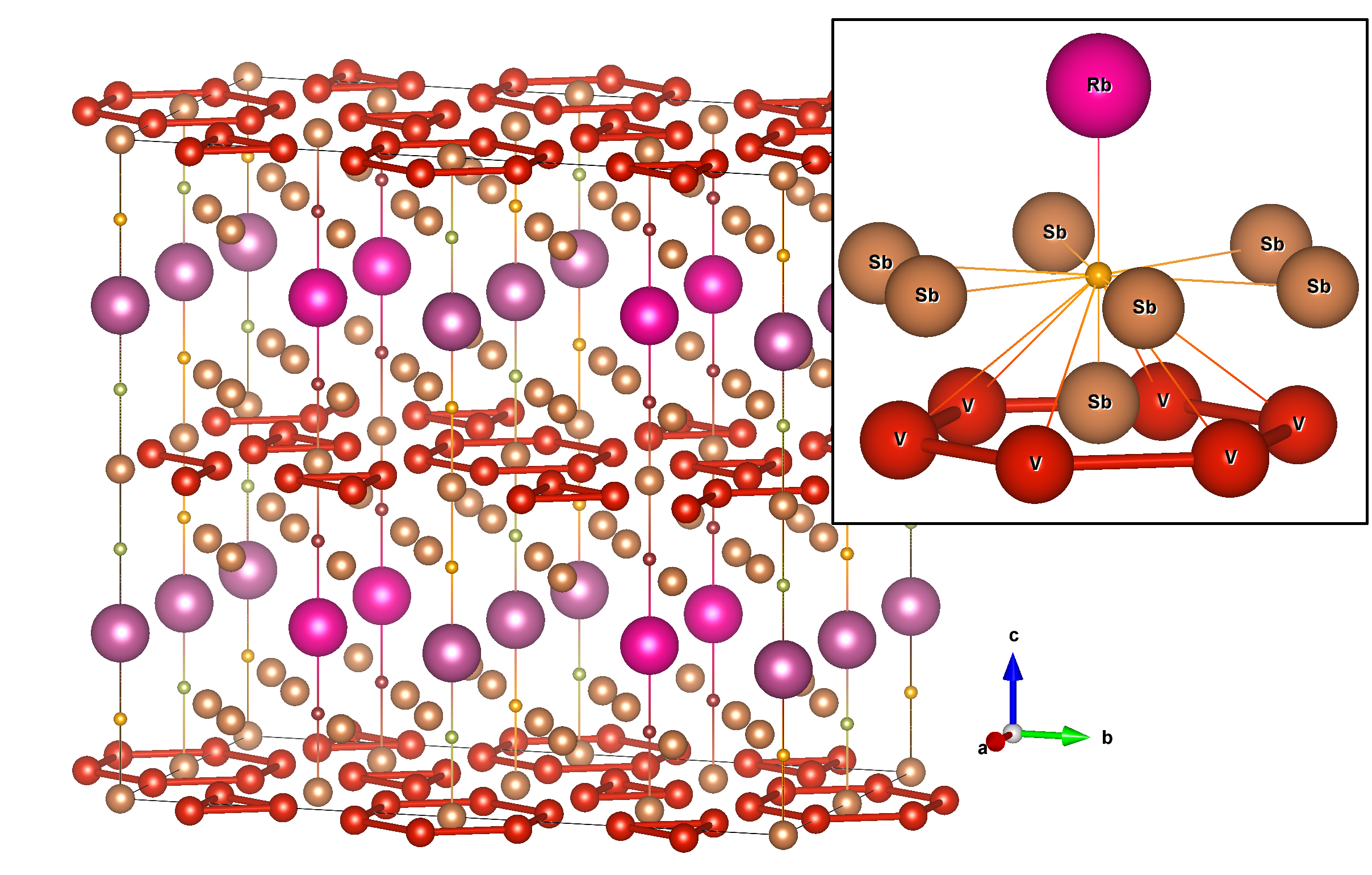}
\caption{The stable muon sites in \rvb obtained with the DFT+$\mu$ method. The orange, green, and dark red spheres indicate stable muon sites, while other atoms are labeled in the inset, where a zoom is shown.
Thin lines indicate the various neighbors of the muon.
The inset also shows the displacement of nearest neighbor Sb atom from the kagome plane formed by V atoms.
The details are provided in Supplementary Table~\ref{tab:muons}. \label{fig:muons}}
\end{figure*}

Three symmetry inequivalent stable muon sites are obtained in \rvb.{}
They are reported in Supplementary Table~\ref{tab:muons}
with labels A, B and C together with atomic positions. 
These sites correspond to the same position in the high temperature hexagonal lattice,
and the respective local neighborhoods are negligibly different (see column four of Supplementary Table~\ref{tab:muons}).
Therefore, we will refer to sites A, B and C in Supplementary Table~\ref{tab:muons} as the single muon site present in the system.

As already mentioned, the structural relaxation yields other interstitial positions that are either high energy sites or are found to be unstable owing to the small energy barriers (smaller than $\sim 0.5$~eV, the average value for the zero point motion energy of the muon) separating them from the stable site already discussed.
The full set of Nudged Elastic Band (NEB) performed to find minimum-energy paths and energy barriers for muon migration in \rvb{} is available in the online repository at https://archive.materialscloud.org/record/2024.30.

%\begin{tabular}{| c | c | c | c | c | c | c |}

\begin{table*}[]
\begin{tabular}{l|c|c|c}
Label (Wyckoff position)    & $E-E_{A}$ (eV) & Coordinates & NN (\AA) / NNN (\AA) / ...  \\
\hline
A (8i)        & 0    &  (0,0,0.121)     & Sb(1.76) Rb(2.66) 6Sb(3.15) 6V(3.56-3.57) \\
B (8i)        & 0.14 &  (0,0,0.380)     & Sb(1.77) Rb(2.61) 6Sb(3.15) 6V(3.52-3.54) \\
C (16j)        & 0.07 &  (0.25,0.25,0.119) &  Sb(1.75) Rb(2.60) 6Sb(3.14) 6V(3.52-3.55)        
\end{tabular}
\caption{Stable muon sites in \rvb.
The second column indicates the energy difference of each site with respect to the lowest energy site \textbf{A}.
The positions, reported in the third column, are given according
to the $Fmmm$ lattice structure described in the text.
The last columns reports the atoms of the various shells of neighbors
and their distance from the muon.
 \label{tab:muons}}
\end{table*}

The EFGs of the nuclei far from the muon 
compare reasonably well with the experiment \cite{Frassineti2023}
as well as with previous simulations (see archive at \cite{Frassineti2023Archive} for further details).
Close to the muon site, the EFG of the nearest neighboring Sb atom is
strongly perturbed and $V_{zz}$ is reduced by $\sim 60$\%.

\subsubsection{Zero point energy}
It is well known that the small mass of the muon implies that zero-point
motion (ZPM) effects are not negligible,
and anharmonicity may play a key role \cite{Onuorah2019,Gomilsek2023,Deng2023}.
This is the case also for the present analysis.
Given the large mass of the neighboring Rb and Sb nuclei,
the potential felt by the muon can be approximated with the
``double Born Oppenheimer approximation''.
A noticeable departure from the harmonic behavior is observed especially along the \emph{c} direction,
as shown in Supplementary Figure~\ref{fig:ana}.
In order to take this into account, the average value of the position operator of the muon is obtained as a function
of the temperature as
\begin{equation}
\langle z\rangle=\frac{1}{Z} \sum_{i} e^{-\beta E_{i}}\left\langle\psi_{i}|z| \psi_{i}\right\rangle \equiv \frac{1}{Z} \sum_{i} e^{-\beta E_{i}} \int dz \, z\left|\psi_{i}(z)\right|^{2}    \label{eq:pos}
\end{equation}
where $Z=\sum_{i} e^{-\beta E_{i}}$ and $E_{i}$ are the solution of the
one-dimensional Schr\"odinger equation $\mathcal{H}\psi_{i}=E_{i} \psi_{i}$.
The Hamiltonian describing the muon is obtained by fitting the potential energy surface obtained ab-initio with a 4-th order
polinomial expansion that gives
\begin{equation}
    \mathcal{H} = \mathcal{T} + \mathcal{V}(z) = \mathcal{T} + a z^{4} + bz^{3} + cz^{2}
\end{equation}
with $\mathcal{T}$ being the kinetic energy operator and $a,b,c$ being
5.57~eV/\AA$^4$ ,-7.59 eV/\AA$^3$ and 5.28 eV/\AA$^2$ respectively.

The inset of Supplementary Figure~\ref{fig:ana} shows the displacement with respect to the equilibrium position obtained from DFT of the expectation value
of the muon position along $z$ obtained with the average in Eq.~\ref{eq:pos} extended over the 20 lowest energy eigenstates.
A shift of 0.062~\AA{} away from the in-plane Sb atom is observed (see Supplementary Figure 1(b)). This alters
the expected depolarization rate substantially, as will be shown in the next sections.
Anharmonicity is at the origin of thermal expansion phenomena and may therefore
contribute to a variation of the muon position as a function of T,  thereby influencing the depolarization rate.
Nevertheless, this effect is negligible within the temperature range of interest and is only relevant above 1000~K.

\begin{figure}
\includegraphics{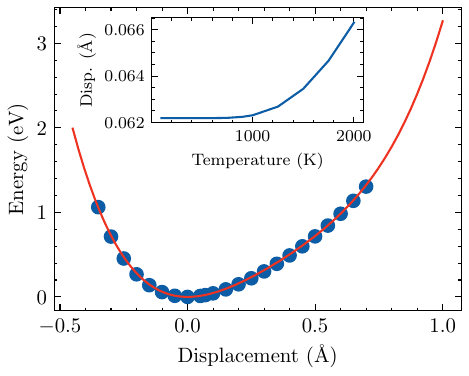}\caption{The potential felt by the muon as it is displaced from the equilibrium position A along the $c$ direction.
Negative(positive) values bring the muon closer to Sb(Rb). The inset shows the expectation value
for the position operator $\langle z \rangle$ (see main text) of the muon along the $z \parallel c$ direction.  \label{fig:ana}}
\end{figure}

\subsubsection{Prediction of zero field signal}

In the absence of magnetic moments of electronic origin, the dipolar interaction with the nuclear moments surrounding the muon assumes a central role.
Accurate predictions can be obtained in this case entirely \emph{ab initio} using
DFT once the perturbation introduced by the muon
on the position and the EFG of the surrounding nuclei
is considered \cite{Sanna2022}.

The muon polarization function \cite{Yaouanc}
can be computed with various
approaches that have been already described in the literature \cite{MCelio, DBillington}.
In this work we adopt the method proposed by Celio \cite{MCelio}
to compute the time evolution of the polarization function of the muon
interacting with a set of (neighboring) $I>1/2$ nuclei, with $I$ being the nuclear spin,
all subject to an electric field gradient (EFG) through the quadrupolar interaction.
The simulations have been performed with the UNDI code \cite{Bonfa2021} and using different sets of
neighbors of the muon that we refer to as ``clusters'' in the following.

%Since the muon polarization undergoes a small change at $T^* < T_{CDW}$, in the next sections
%we describe in details all the approximations that have been used to estimate its trend from first principles.

\subsubsection{Exponential growth}
The dimension of Hilbert space grows exponentially with the number of nuclei included
in the simulation. The limit of contemporary computational resources is rapidly reached when attempting to converge the predicted muon polarization as a function of the number of nearest neighbors considered in the simulation.
A simple approximation has been proposed by Wilkinson and Blundell \cite{Wilkinson}
to sidestep the problem:
The contribution of the remaining nuclei in the system can be effectively included by
enhancing the dipolar interaction of certain nuclei within the simulation,
typically those situated farthest from the muon.
The scaling is defined in such a way that the
contribution to the second moment of the nuclear dipole field distribution
from the set of nuclei included in the simulations equals the one of the entire system.
This approach has a few drawbacks.
The evaluation of the second moments neglects the quadrupolar interaction, for which an analytical
expression is not generally possible \cite{EFG}.
In addition, the scaling alters the ratio between the Zeeman and the quadrupolar
interactions of the nuclei under consideration.
For this reason, the accuracy of this approximation, dubbed \emph{effective rescale approximation} (ERA), in carefully checked in this section.

\begin{figure}
\includegraphics{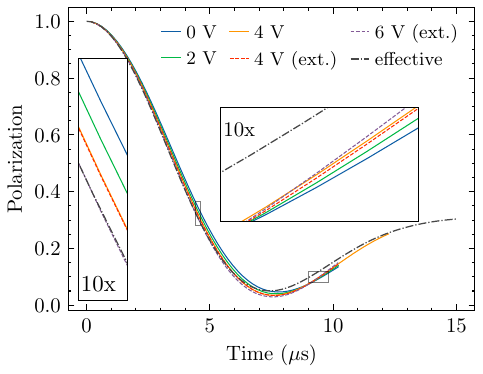}
\caption{Muon spin polarization for the \ce{^{85}Rb^{121}Sb7 ^{51}V_{x}} cluster as a function of the
         number $x$ of V atoms included in the simulation.
         The continuous lines are exact results while the dashed lines (labeled ext) are extrapolated from the 2 V results (see main text).
         The dash-dotted line is the prediction obtained with the ERA applied to the six
         $^{121}$Sb third nearest neighbors of the muon to account for the next nearest neighboring V nuclei. \label{fig:checkwb}}
\end{figure}

Supplementary Figure~\ref{fig:checkwb} shows the exact trends obtained for the clusters
\ce{^{85}Rb^{121}Sb7 ^{51}V_{x}} with x=0,2,4 obtained by progressively
including couples of opposite V nuclei composing the hexagon of Supplementary Figure~\ref{fig:muons}.
A simple extrapolation strategy can be used to obtain the x=6 trend: assuming that
$P^{(\ce{RbSb7V2})}(t) = P^{(\ce{RbSb7})}(t) \exp \left[ (\sigma^{(\ce{V2})} t) ^{2} \right]$
from the ratio $P^{(\ce{RbSb7V2})} / P^{(\ce{RbSb7})}$ the exponential relaxation
produced by the \ce{V2} couple can be extracted. The value obtained for $\sigma^{(\ce{V2})}$ is proportionally increased to account for the remaining V nuclei.
This method predicts correctly the exact  trend for \ce{V4} (continuous orange line and dashed red line in Supplementary Figure~\ref{fig:checkwb})
in the entire time interval of interest.

The results obtained with the two approaches
are finally compared in Supplementary Figure~\ref{fig:checkwb} where the full
lines are obtained with exact simulations, dashed lines are extrapolated
and the dash-dotted line shows the ERA
accounting for the 6 nearest neighbors V nuclei in the \ce{^{85}Rb ^{121}Sb7} cluster by rescaling the dipolar interaction of the 6 third nearest neighboring Sb nuclei.
The left inset shows that, at short time, the
relaxation rate for $x=6$ obtained with the two approaches align very well.
On the contrary, a small deviation is observed for $t>$~6~${\mu}s$
which however remains smaller than 5\% throughout the entire relevant time span.

\subsubsection{Isotope average}

A simple way to obtain the isotopic forms of a cluster is achieved through
the symbolic expansion of a polynomial function  \cite{Yamamoto,Brownawell}.
This is of interest when performing isotopic averages for a selected set of
nuclei.
For example, when considering the cluster \ce{RbSb7}, the polynomial takes the form
\begin{equation}
    ({}^{85}\text{Rb} + {}^{87}\text{Rb})^{1} \times ({}^{121}\text{Sb} + {}^{123}\text{Sb})^{7} \label{eq:isotopes}
\end{equation}
The symbolic expansion of Eq.~\ref{eq:isotopes} yields numerous product terms,
each representing distinct isotopic variants of the cluster.
Substituting the isotopic abundances for Rb and Sb
into each term allows for the determination of the probability of each cluster.
The accuracy of the isotope average can now be progressively improved
by including different structures in decreasing order of probability.
The results for the \ce{RbSb7} cluster are shown in Supplementary Figure~\ref{fig:isoavg1}.
Yet this strategy can be very time-consuming since all different
permutations in the position of the isotopes must be considered \cite{Symmetry}.

\begin{figure}
\includegraphics{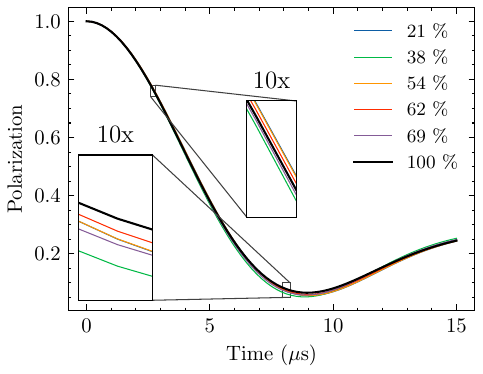}
\caption{Muon spin polarization for the \ce{RbSb7} cluster as a function of the
         isotopic distributions considered in the isotope average. The legend shows the cumulative probability
         covered by the selected set of structures. \label{fig:isoavg1}}
\end{figure}

\begin{figure}
\includegraphics{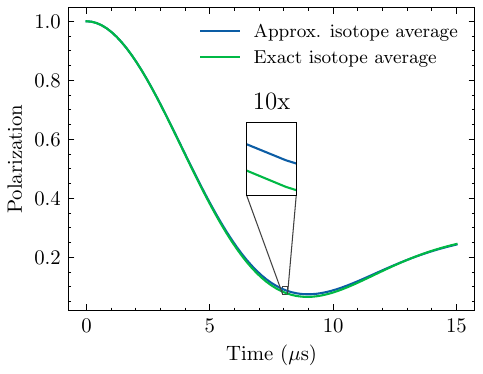}
\caption{Comparison between the exact and the approximate (see text) isotope average for the \ce{RbSb7} cluster. \label{fig:isoavg2}}
\end{figure}

An alternative approximate approach involves considering only those configurations
where all nuclei are of one isotope kind, and then averaging the results based on the
product of the abundances for each isotope appearing in the simulation.
For the case of \ce{RbSb7}, this results in four simulations.
The accuracy of this approximation is illustrated by Supplementary Figure~\ref{fig:isoavg2}.

While the exact isotope average for the
\ce{RbSb7} cluster can be easily obtained, this latter approach makes the simulations
much faster at a negligible accuracy loss and, for this reason,
we adopt this strategy in the discussion that follows.

\subsection{Comparison with experiment}

\subsubsection{Effect of CDW transition}
The kagome lattice formed by V atoms in \rvb undergoes the most significant
distortion when the CDW order sets in.
This transition is barely observed in the experiments, and the reason is now easily understood.
The V atoms are more than 3.5~\AA{} far from the muon.
The dipolar nature of the interaction makes the displacement of V atoms in the kagome plane
occurring at $T_{CDW}$
barely visible in the \msr signal in light of the large distance $d_{V-\mu}$ between V atoms and the muon.
In order to estimate an upper bound for this effect, we computed the
variation of the muon relaxation considering a reduction of the
distance $d_{V-\mu}$ of 0.17~\AA{} in the
\ce{RbSbV6} cluster.
This is a factor $2\div3$ larger than the estimated
displacement of the V atoms from their position
in the pristine hexagonal lattice after the CDW
sets in \cite{Kautzsch,ScagnoliGuguchia}. The expected trends are shown in Supplementary Figure~\ref{fig:v} while the experimental asymmetries acquired above and below $T_{CDW}$, respectively at 120~K and 75~K, are shown in Supplementary Figure~\ref{fig:exp} and almost perfectly overlap.

\begin{figure}
\includegraphics{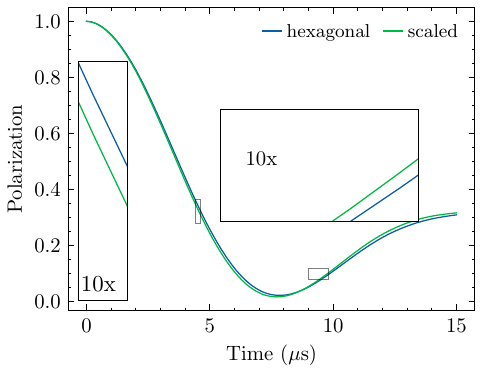}
\caption{Expected depolarization rate for the \ce{RbSbV6} cluster with distance $d_{V-\mu}$
changing from 3.57~\AA{} to 3.40~\AA{}. \label{fig:v}}
\end{figure}

\subsubsection{Experimental vs ab initio muon polarization}
The \emph{ab initio} prediction for the muon polarization
in the absence of electronic moments is shown by the red
line in Supplementary Figure~\ref{fig:exp}.
This result is obtained with the \ce{RbSb7} cluster using the ERA to
include the contribution of all the remaining nuclei of the compound.
In light of the previous discussion, the error due to this approximation is expected to
be negligible up to 6${\mu}s$ (see Figs.~\ref{fig:checkwb}).
The \emph{ab initio} prediction appears to be slightly faster than the experimental signal.
A displacement of +0.03~\AA{} from the predicted equilibrium position is sufficient for restoring a very good agreement. This quantity is less than half of the Zero-Point Motion correction, thus emphasizing the significance of this latter contribution.

\begin{figure}
\includegraphics{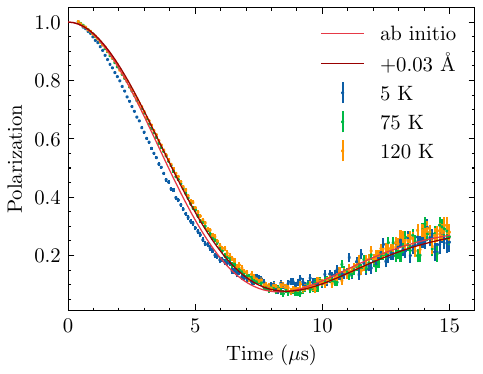}
\caption{
Zero-field \msr measurements of \rvb at three temperatures. The red solid curve
shows the ab initio prediction in the absence of electronic moments.
The brown line is obtained by displacing the muon from its equilibrium position
by only 0.03~\AA{} in order to restore almost perfect agreement with the experimental results. \label{fig:exp}}
\end{figure}


\begin{thebibliography}{150}
\bibitem{Footnote} This estimate does not include the uncertainties related to the ab initio estimates of atomic positions and electric field gradients.
\end{thebibliography}
\begin{thebibliography}{150}
\bibitem{Symmetry} Point symmetry can be used to reduce the computational effort to some extent.
\end{thebibliography}
\end{document}